\documentclass[a4paper,12pt]{article}
\usepackage[utf8]{inputenc}
\usepackage{amsfonts,amsmath}

\usepackage{color}
\usepackage{tabularx}

\usepackage{dsfont}
\usepackage{times}

\usepackage{ifthen}
\usepackage{enumerate}
\usepackage{amsmath}
\usepackage{amssymb}
\usepackage[mathscr]{eucal}
\usepackage[errorshow]{tracefnt}
\usepackage{amsmath}
\usepackage{slashed}
\usepackage{amssymb}
\usepackage{array}
\usepackage{amsthm}
\usepackage{eucal}
\usepackage{epsf}
\usepackage{epsfig}
\usepackage{psfrag}
\usepackage{multirow}
\usepackage{pdfsync}

\def\Tr{{\rm Tr} \,}
\def\pmb#1{\setbox0=\hbox{#1}
\kern.0em\copy0\kern-\wd0 
\kern-.04em\copy0\kern-\wd0 
\kern.08em\copy0\kern-\wd0 
\kern-.04em\raise.0433em\box0 }         
 
\newcommand{\nc}{\newcommand} 
\nc{\beq}{\begin{equation}} 
\nc{\eeq}{\end{equation}} 
\nc{\ber}{\begin{eqnarray}} 
\nc\eer[1]{\label{#1}\end{eqnarray}} 
\newcommand{\bea}{\begin{eqnarray}} 
\newcommand\eea{\end{eqnarray}}

\newcommand{\CN}{\mathcal{N}}
\def\be{\begin{equation}} 
\def\ee{\end{equation}}
\nc{\pek}[1]{\cite{#1}} 
\nc{\enr}[1]{(\ref{#1})} 
\nc{\kal}[1]{{\cal{#1}}} 
\nc{\dott}{\;\cdot\;} 

\def\0 {\nonumber}

\def\tq{{\widetilde q}}
\def\tM{\widetilde{M}}
\def\tr{{\rm Tr}\,}

\begin{document} 


\thispagestyle{empty}
\begin{flushright}
IFUP-TH/2015
\end{flushright}
\vspace{10mm}
\begin{center}
{\Large     ${\cal N}=2$ Argyres-Douglas theories, \\
${\cal N}=1$ SQCD and Seiberg duality} 
\\[15mm]
{Stefano Bolognesi$^{a,b}$,   Simone Giacomelli$^{c}$,   Kenichi Konishi$^{a,b}$  
} \footnote{\it e-mails: ~stefanobolo(at)gmail.com,  ~ simone.giacomelli(at)ulb.ac.be, ~
konishi(at)df.unipi.it}
\vskip 6mm
 
\bigskip\bigskip
{\it  
$^a$  
~Department of Physics ``E. Fermi'', University of Pisa, \\
Largo Pontecorvo, 3, Ed. C, 56127 Pisa, Italy
\\
$^b$
  INFN, Sezione di Pisa,
Largo Pontecorvo, 3, Ed. C, 56127 Pisa, Italy 
\\
$^c$
Universit\'e  Libre de Bruxelles and International Solvay Institutes,  \\
ULB-Campus Plaine CP231 1050 Brussels, Belgium.
\\
  }
\vskip 6 mm

\bigskip
\bigskip

{\bf Abstract}\\[5mm]
{\parbox{14cm}{\hspace{5mm}
\small

We revisit the study of singular points in the Coulomb branch of $\mathcal{N}=2$ SQCD in four dimensions with gauge group $SU(N)$. 
For certain choices of the mass parameters these vacua are not lifted by a mass term for the chiral 
multiplet in the adjoint representation. By using recent results about the M5 brane description of $\mathcal{N}=1$ theories we 
study the resulting vacua and argue that the low-energy effective theory has a simple Lagrangian description involving a free chiral multiplet in the adjoint representation of the flavor symmetry group, 
a system somewhat reminiscent of the standard low-energy pion description of  the real-world QCD. This fact is quite  remarkable in view of the fact that the underlying  $\mathcal{N}=2$ SCFT  (the Argyres-Douglas systems) are strongly-coupled  non-local theories of quarks and monopoles. 
}
}
\end{center}
\newpage
\pagenumbering{arabic}
\setcounter{page}{1}
\setcounter{footnote}{0}
\renewcommand{\thefootnote}{\arabic{footnote}}

\tableofcontents

\section{Introduction}

In the past few years several techniques have been developed to understand the properties of field theories without a Lagrangian 
description. A well-known example is the M-theory realization 
of $\mathcal{N}=2$ theories in four dimensions proposed some years ago by Gaiotto \cite{gaiotto}. 
More recently this construction has been extended to a large class of four dimensional theories with $\mathcal{N}=1$ 
supersymmetry and, as in the $\mathcal{N}=2$ case, the information about the protected sector of the theory is encoded in an 
auxiliary complex curve \cite{noi,xie1,xie2,kazuya,IS3}. In the M-theory description this curve arises as a holomorphic cycle in a Calabi-Yau 
threefold and in the case of superconformal theories the R-charge (and hence the dimension) of chiral operators can be read from 
the curve imposing that the holomorphic three-form (which encodes the low-energy effective superpotential \cite{witten}) 
has R-charge two \cite{ionuovo}. Using these tools one can for example study RG flows between nonLagrangian $\mathcal{N}=1$ 
theories (provided the corresponding curve is known) and read out properties of the IR fixed point. 

In this paper we use this idea to explore the infrared behaviour of $\mathcal{N}=2$ SQCD softly broken by a mass term for the 
chiral multiplet in the adjoint representation of the gauge group: for generic values of the parameters the 
low-energy effective theory at the points in the $\mathcal{N}=2$ Coulomb branch which are not lifted by the superpotential has a 
simple Lagrangian description. Indeed, just from the knowledge of the Seiberg-Witten (SW) solution \cite{SW1,SW2}, it is possible to understand the 
effect of the superpotential deformation. In this way one can understand confinement and chiral symmetry breaking as a consequence
of monopole condensation \cite{SW1}-\cite{Carlino:2000uk}. However, for certain choices of the parameters the effective theory 
at the relevant points in the Coulomb branch is a strongly coupled theory with relatively nonlocal degrees of freedom, analogous 
to the theory discovered by Argyres and Douglas \cite{AD} (AD theory). Such theories do not have any Lagrangian description and, consequently,  it appears to be a rather formidable task  to find out  the effect of the superpotential and hence the properties of the theory in the IR, from the SW solution 
alone \footnote{In some systems of this kind, but in theories with $N_f$ even, a dual description recently found by Gaiotto, Seiberg and Tachikawa \cite{GST}   can be used to analyze the phases of the far-infrared systems.  See \cite{GK} and references cited therein.}.     As we shall see, by using 
the methods mentioned above one can determine the R-symmetry preserved by the $\mathcal{N}=1$ deformation 
and consequently determine several properties of the resulting IR fixed point exploiting the results of \cite{TS}. 

Our methods are general and can be applied for generic choices of the superpotential breaking the ${\cal N}=2$ supersymmetry. The special feature of the quadratic 
superpotential  (on which we focus here) is that the resulting IR effective theory turns out to be extremely simple: after the $\mathcal{N}=1$ 
deformation the AD theory flows to a theory describing a free chiral multiplet in the adjoint representation of the global symmetry 
group of the underlying gauge theory.  Small variations of the parameters of the gauge theory result in some superpotential 
terms for the chiral multiplet. One thus finds an effective low-energy description in terms of gauge invariant degrees of freedom, 
which is quite reminiscent of  the effective pion Lagrangians in the real-world QCD. Of course, in the supersymmetric setup we are working in, there is much better  control than in QCD and one can actually derive precisely the form of the Lagrangian from the underlying microscopic theory.

As the physical effect of adding an adjoint scalar mass is to make the system more strongly coupled in the infrared, it is 
perhaps reasonable to interpret the free mesons found here as composites of the dyons of the ${\cal N}=2$ AD theories.  

The rest of the paper is organized as follows: in Section 2 some known facts about $\mathcal{N}=2$ SQCD and the strongly 
coupled theories of interest for us are reviewed. In section 3 we study the effect of the superpotential deformation and give evidence that 
the resulting IR fixed point is a free theory by computing the a and c central charges. As a nontrivial consistency check we 
show that our candidate IR effective theory satisfies 't Hooft anomaly matching condition. In Section 4 we study the moduli space of the 
gauge theory and check that the F-term equations arising from the superpotential of our effective theory reproduce in detail 
the results of the gauge theory computation. This can also be seen as a further consistency check of the techniques developed in 
\cite{noi,ionuovo}. In  Appendix A we re-derive the results of Section~\ref{sec:GTP} and argue that the chiral multiplet in the adjoint of the flavor group found in the first sections  can be identified with 
the meson appearing in Seiberg's dual description of SQCD \cite{Sei}.

\section{Singular points on the Coulomb branch of $\CN=2$ SQCD  \label{sec:Singular}} 

In this section we describe the superconformal theories we are interested in and review their main properties. The content of 
this section is mainly based on \cite{APSW} and \cite{EHIY}. 

\subsection{$SU(2)$ with  $N_f=1$}

The SW curve can be written in the form \cite{witten2}
\begin{equation}\Lambda(v+m)t+v^2+u+\frac{\Lambda^2}{t}=0\;;\qquad \lambda=\frac{v}{t}dt\;,\end{equation}
or equivalently, after a redefinition of $v$, as 
\begin{equation}v^2=\frac{\Lambda^2}{4t}(t^3-4\frac{m}{\Lambda}t^2-4\frac{u}{\Lambda^2}t-4)\;.\end{equation}
By tuning $m$ and $u$ appropriately the curve becomes singular and reduces to 
\begin{equation}v^2=\frac{\Lambda^2}{4t}(t-4^{1/3})^3.\end{equation}
With the change of variable $z=t-4^{1/3}$ and after a suitable scaling limit (see e.g. \cite{APSW}) this becomes 
\begin{equation}v^2 = z^3\;;\qquad \lambda=vdz\;.\end{equation}
These are the curve and differential for the AD theory of type $A_2$. This theory is known to have a coupling constant, 
which we call $m$, of dimension $4/5$ and a chiral operator $U$ of dimension $6/5$. When we embed the theory in $SU(2)$ SQCD with one flavor as we have 
done above, the coupling constant and chiral operator of the SCFT are inherited from the mass parameter and Coulomb branch operator 
of the gauge theory respectively. A change of the value of the mass parameter in the gauge theory  by  \footnote{Throughout,  we shall  denote by $\delta m$  the 
common shift of the quark masses from the critical value, $m=m^*$ at the AD point.   The symbols ${\tilde m}_i$   ($\sum_i  {\tilde m}_i=0$),  will be   instead reserved for the nonAbelian flavor dependent deviation of the masses. }   $\delta m$
is mapped in the IR to the following $\CN=2$ preserving deformation \cite{APSW}
 \begin{equation}\int d^2\theta d^2\tilde{\theta} \, \delta m\, U\;.\end{equation}
In the following we shall need  $a$ and $c$ central charges of this theory, which were computed in \cite{TS}. The result is 
\begin{equation}   a=\frac{43}{120}\;;\qquad c=\frac{11}{30}\;.  \end{equation}

\subsection{$SU(2)$ with $N_f=2$}

This is the only model with even number of flavors we consider. The SW curve can be written as 
$$\Lambda^2t+v^2+u+\frac{(v+m_1)(v+m_2)}{t}=0\;;\qquad \lambda=\frac{v}{t}dt\;.$$ 
With a change of variable we can bring it to a Gaiotto inspired form 
$$v^2=\frac{-\Lambda^2t^3-(\Lambda^2+u)t^2-(u+m_1m_2)t+(m_1-m_2)^2/4}{(t+1)^2}\;;\qquad \lambda=v\frac{dt}{t}.$$ 
Setting $u=-\Lambda^2$, $m_1=m_2=\Lambda \equiv m^*$ and performing the scaling limit we find 
$$v^2=t^3\;;\qquad\lambda=v\frac{dt}{t}\;.$$
 We recognize here the SW curve for the AD theory of type $D_3$, which is the same as the AD theory 
of type $A_3$. This theory is known to have a $SU(2)$ global symmetry, which in the present construction is inherited from the 
global symmetry of the gauge theory. 

As every $\mathcal{N}=2$ SCFT, our theory includes a chiral operator transforming in the adjoint 
of the global symmetry (the lowest component of the multiplet containing the conserved current) which is commonly called moment map (see 
Fact 2.1 of \cite{tienne} and \cite{index} for further details).   The corresponding chiral multiplet will be denoted as $M_{SU(2)}$. In the present case (as in 
all the other models we are going to  discuss in this note) this is naturally associated with the traceless part of the 
meson $\widetilde{Q}Q$ of the gauge theory. The a, c and $SU(2)$ central charges are \cite{TS} 
$$ a=\frac{11}{24}\;;\qquad c=\frac{1}{2}\;;\qquad k_{SU(2)}=\frac{8}{3}\;.$$

If we wish to slightly change the value of the mass parameters and understand how this ``deformation'' is mapped in the AD theory, we should 
distinguish two cases: setting $m_1=\Lambda+ {\tilde m}$; $m_2=\Lambda-  {\tilde m}$ we break the $SU(2)$ global symmetry to the Cartan $U(1)$. 
Because of the above-mentioned relation between the moment map of the AD theory and the meson of the gauge theory, we have a natural 
candidate deformation in the AD theory: the superpotential term
 $$\mathcal{W}=\int d^2\theta \, {\tilde m}\,\Tr(\sigma_3M_{SU(2)})\,.$$ 
Setting instead $m_1=m_2=\Lambda+\delta m$, which does not break the $SU(2)$ global symmetry, corresponds as in the $N_f=1$ case to 
turning on the $\CN=2$ preserving term \cite{APSW}
 $$\int d^2\theta \, \delta m \,V;\qquad V=\int d^2\tilde{\theta}\, U\;.$$ 
 In the above 
formula $U$ is the dimension $4/3$ Coulomb branch operator of the AD theory and $\delta m$ is interpreted as the corresponding 
dimension $2/3$ coupling constant. 
The generic mass deformation is just a combination of the two special cases considered above and is consequently equivalent to 
activating both superpotential terms. 

\subsection{$SU(2)$ with three flavors}

The SW curve is 
$$t\Lambda(v+m_1-\Lambda/2)+v^2+u+\frac{(v+m_2-\Lambda/2)(v+m_3-\Lambda/2)}{t}=0\;;\quad \lambda=v\frac{dt}{t}\;.$$
First of all we bring it to the 6d inspired form (with the same SW differential)
$$v^2=\frac{\left[t^2\frac{\Lambda}{2}+\left(\frac{m_2+m_3}{2}-\frac{\Lambda}{2}\right)\right]^2}{(t+1)^2}- 
\frac{t^2\Lambda(m_1-\Lambda_2)+ut+(m_2-\Lambda/2)(m_3-\Lambda/2)}{t+1}\;.$$ 
The singular point we are after is found  by setting $u=0$ and $m_1=m_2=m_3= \Lambda/2 \equiv m^*$. Performing the scaling limit we find 
$$v^2=t^4\;;\qquad \lambda=v\frac{dt}{t}\;,$$
 which is the SW curve for the AD theory of type $D_4$. This theory has an $SU(3)$ 
global symmetry inherited from  the underlying gauge theory. We have an operator $U$ of dimension $3/2$ 
and the corresponding dimension $1/2$ coupling which can be identified with the $U(1)$ mass parameter of the gauge theory \cite{APSW} 
$\delta m\equiv(m_1+m_2+m_3)/3-\Lambda/2$. An argument analogous to the one given in the previous subsection tells us that the other 
two mass deformations are mapped to the following superpotential terms in the AD 
theory: 
$$\int d^2\theta\,  {\tilde m}_3 \,\Tr(T_3M_{SU(3)})\;;\qquad \int d^2\theta \,  {\tilde m}_8 \, \Tr(T_8M_{SU(3)})\;,$$
 where $M_{SU(3)}$ is the $SU(3)$ 
moment map of the AD theory and $T_3$, $T_8$ denote the Cartan generators of $SU(3)$ in the fundamental representation.

\subsection{$SU(N)$ theories \label{sec1:SUN}} 

Let us  move now to $SU(N)$ SQCD, limiting ourselves to the particular case with $N_{f}=2N-1$ flavors. 
The properties of the superconformal theories we are going to discuss depend only on $N_f$, so this is actually not restrictive. 
The case of $SU(2)$ SQCD with three flavors discussed in the previous section 
falls in this class. It is convenient to write the curve in the form \cite{APS}
$$y^2=P_N(x)^2-4\Lambda\prod_i\left(x+m_i-\frac{\Lambda}{N}\right).$$ 
By setting $m_i=m^*=  \Lambda/N$ and all the $u_i$ to zero the curve degenerates to  
\begin{equation}\label{ciao}y^2=(x-4\Lambda)\, x^{2N-1}\;, \end{equation} 
and after the scaling limit we are left with the curve 
\begin{equation}y^2=x^{2N-1};\qquad \lambda_{SW}=x\frac{dy}{y}\;.\end{equation} 
These superconformal theories have already been studied in \cite{EHIY} (they also appear in \cite{iocecotti} where they were 
called $D_2(SU(2N-1))$). They have 
chiral operators $U_k$ of dimension 
$$D(U_k)=\frac{2k-1}{2}\;;\quad k=2,\dots,N\;,$$ which are inherited from the Coulomb branch operators 
of the parent gauge theory and a coupling constant $m$ of dimension $1/2$ which descends, as in the previous cases, from the 
singlet mass parameter of the gauge theory 
$$\delta m=\frac{1}{2N-1}\sum_im_i-    m^*, \qquad    m^*  \equiv \frac{\Lambda}{N}\;.$$ 
Another important property for our 
analysis is the fact that these models have an $SU(2N-1)$ global symmetry. The a, c and $SU(2N-1)$ central charges 
can be computed by using e.g. the methods of \cite{TS}. The result is (see \cite{iocecotti}) 
\begin{equation}\label{ccc} a=\frac{7}{24}N(N-1)\;;\qquad c=\frac{1}{3}N(N-1)\;;\qquad k_{SU(2N-1)}=2N-1\;.\end{equation} 
As in the previous cases, a variation of the singlet mass parameter gets mapped to the relevant deformation
\begin{equation}\int d^2\theta \, \delta m \, V_2\;; \qquad V_2=\int d^2\tilde{\theta}\, U_2\;,\end{equation} in the superconformal theory. 
Other (flavor non-singlet) mass deformations correspond to activating a superpotential term linear in the $SU(2N-1)$ moment map: 
$$\int d^2\theta \, \Tr\left[\left(\sum_i\tilde{m}_iT_i\right)M_{SU(2N-1)}\right]\;, $$
where  $T_i$ are the Cartan generators of $SU(2N-1)$ 
in the fundamental representation.


\section{$\mathcal{N}=1$ deformation  \label{sec:N=1}} 

We shall  now deform the gauge theories mentioned in the previous section by adding a mass term for the chiral multiplet in the adjoint 
($\mu \int d^{2}\theta \,\Tr\Phi^2$). Of course this superpotential breaks extended supersymmetry and most of the Coulomb branch is lifted;   
however, it turns out that the AD-like points we described before are not lifted. The $\mathcal{N}=2$-breaking term is then mapped 
in the IR to some relevant deformation of the AD theory, which will then flow to some $\mathcal{N}=1$ infrared fixed point. 
The goal of the present section is to better understand the resulting superconformal theories and, as we shall see, the answer turns out to 
be quite surprising. Our tool to address this problem is the $\mathcal{N}=1$ curve studied in \cite{noi} (see also 
\cite{xie1,xie2,kazuya,IS3}).

\subsection{$SU(2)$ theory with one flavor}

Let us recall first of all how to determine the $\mathcal{N}=1$ curve for this theory: we start from the curve for the gauge 
theory with a massive chiral multiplet (of mass $\mu$) in the adjoint representation at the relevant point on the Coulomb branch (see \cite{noi}) 
\begin{equation}\begin{cases} v^2=\frac{\Lambda^2z^3}{4(z+4^{1/3})}\;;\\
 w^2=\frac{\mu^2\Lambda^2}{4}  (z^2+az+b)\;; \\
 w=f(z)v\;, 
\end{cases}  \;\qquad \Omega=\frac{dvdwdz}{z+4^{1/3}}\;. \end{equation} 
The first equation is the $\CN=2$ curve and the second is obtained by imposing the boundary condition $w\sim\mu v$ for large $z$. 
The parameters $a$ and $b$ can be fixed by exploiting the constraint $w=f(z)v$ which tells us that the ratio $w^2/v^2$ has to be the 
square of a rational function on the $z$ plane. This is a consequence of the Hitchin field description and tells us that 
the spectral curve is a two-sheeted covering of the z plane as required by the 6d realization of the theory. In more physical 
terms, it was shown in \cite{noi} that this constraint is equivalent to imposing the factorization condition (see e.g. \cite{DS}) 
which identifies the points on the Coulomb branch which are not lifted by the $\CN=1$ deformation. The constraint leads to $b=0$, $a=4^{1/3}$ and 
consequently $f(z)=\mu(z+4^{1/3})/z$. The space parametrized by $v$, $w$ and $z$ is Calabi-Yau and $\Omega$ is the corresponding 
holomorphic three-form.
As in the case of $\CN=2$ theories, we now need to identify a suitable scaling limit in order to extract the spectral curve 
associated with the $\CN=1$ AD theory. Our proposal is the following: as in the $\mathcal{N}=2$ case we send $\Lambda$ to 
infinity (i.e. we discard terms proportional to a negative power of $\Lambda$) after the redefinition \footnote{the exponents are 
chosen in such a way that $\Omega$ remains finite in this limit.}
\begin{equation}v=\Lambda^{2/5}v'\;; \qquad z=\Lambda^{-2/5}z'\;;\qquad \mu=\Lambda^{-4/5}\mu'\;.\end{equation} 
Implementing this we end up with the curve
\begin{equation}\label{Nf1}\begin{cases} v'^2=z'^3 \;;\\
 w^2=\mu'^2z'\;;\\
 z'w=\mu'v'\;, 
\end{cases} \qquad  \Omega=dv'dwdz' \;. \end{equation} 
Notice that the first equation is simply the SW curve of the AD theory of type $A_2$ and $v'dz'$ is the corresponding SW differential. 
Imposing now the constraint \cite{ionuovo} $D(\Omega)=3$ (or equivalently $R(\Omega)=2$)  we find \footnote{Here we are assuming that $\mu'$ is dimensionless, which is indeed the 
case if the $\CN=1$ deformation leads to a nontrivial IR fixed point. We regard the fact that the final answer is a consistent SCFT 
as a nontrivial consistency check on the correctness of this assumption.}  the following assignment of R-charges: 
$$ R(v')=1,\quad R(z')=\frac{2}{3};\quad R(w)=\frac{1}{3}.$$ 
Exploiting now the fact that $v'$ and $z'$ have respectively charge $6/5$ and $4/5$ under $R_{\CN=2}$ and zero under $I_3$ 
(the Cartan generator of the $SU(2)$ R symmetry), whereas $w$ has charge one under $I_3$ and zero under $R_{\CN=2}$, we conclude 
that the combination of $R_{\CN=2}$ and $I_3$ of the AD theory preserved by the above $\CN=1$ deformation is 
\begin{equation}  \label{nf1} \frac{5}{6}R_{\CN=2}+\frac{1}{3}I_3 \;.\end{equation}
Notice that the above combination is precisely the one preserved by the following $\CN=1$ deformation of the AD theory: 
$$\int d^2\theta\mu'U\;.$$
 This result fits perfectly with our expectations since, as was explained in the previous section, the 
chiral operator $U$ of the AD theory is related to $\Tr\Phi^2$ of the gauge theory. 
Under the assumption that (\ref{nf1}) is the R-symmetry of the IR fixed point, one  can immediately evaluate the a and c central charges 
of our IR fixed point using the well-known formulas \cite{anselmi}
$$a=\frac{3}{32}(3\Tr R^3-\Tr R)\;;\qquad c=\frac{1}{32}(9\Tr R^3-5\Tr R)\;,$$ 
combined with the relations \cite{TS} 
\begin{equation}\label{scft}\Tr R_{\CN=2}^3=\Tr R_{\CN=2}=48(a'-c')
\;;\qquad \Tr R_{\CN=2}I_3^2=4a'-2c'\;,\end{equation}where $a'$ and $c'$ are the central charges of the AD 
theory (here we are using 't Hooft anomaly matching) given in the previous section. The result is 
$$a=\frac{1}{48}\;; \qquad c=\frac{1}{24}\;,$$
 which are the $a$ and $c$ central charges of a free chiral multiplet.
We are thus led to the proposal that our system at low energies can be described by a single chiral multiplet which we call $\Psi$. 
The superpotential term $\mu \Tr\Phi^2$ in the gauge theory is then 
mapped to $\mu'\Psi^3$ in the effective low-energy theory (just because its R-charge is three times that of $\Psi$). 

As we have seen in the previous section, changing the mass of the doublet in the gauge theory corresponds in the AD theory to the superpotential 
deformation 
\begin{equation}\int d^2\theta \, \delta m\, V\;;\quad V=\int d^2\tilde{\theta} \, U \;.\end{equation}
Using the fact that $\tilde{\theta}$ has charges $1$ and $-1/2$ under $R_{\CN=2}$ and $I_3$ respectively, we conclude that $V$ has 
charge $2/3$ under (\ref{nf1}), or equivalently dimension one at the IR fixed point. This suggests the identification 
$$\delta m\, V\approx \delta m\, \Psi\;.$$ 
We thus conclude that when the mass of the doublet is slightly different from the critical value, the effect of 
the $\CN=2$ breaking deformation can be effectively described by the superpotential term 
\begin{equation}\mathcal{W}=\delta m\, \Psi+\mu' \, \Psi^3\;.  \label{This}\end{equation}
$\mathcal{W}$  has  two critical points. In each one of the two vacua the second 
derivative of the superpotential is different from zero and is of the order $\sqrt{\delta m \,\mu'}$; as a consequence our chiral multiplet becomes 
massive in both vacua. 

This matches precisely the gauge theory expectation.  As is well known, there are three singular points in the quantum moduli 
space of the ${\cal N}=2$
$SU(2)$ theory with $N_f=1$, where either a quark or a dyon becomes massless \cite{SW2}. The AD point arises when the quark vacuum  and a dyon vacuum 
collide by a judicious choice of the quark mass \cite{APSW}.   Vice versa,  when the quark mass is taken slightly off the critical value, the AD vacuum 
splits into two vacua. These can be analyzed directly by  using the SW solution and are found to develop mass gap. 

\subsection{$SU(2)$ theory with three flavors}

Consider  now  the theory with three flavors. The analysis in this case is similar to the previous one.  
The study of the theory with two flavors requires some particular consideration and will be postponed to the next subsection.

After the $\CN=1$ breaking we find the following spectral curve 
\begin{equation}\begin{cases} v^2=\frac{\Lambda^2t^4}{4(t+1)^2}\;;\\
 w^2=\mu^2\Lambda^2t^2/4\;;\\
 w=\mu(1+1/t)v\;,  
\end{cases}\qquad   \Omega=dvdw\frac{dt}{t}\;. \end{equation} 
In the above formula we gave directly the solution of the constraint $w=f(t)v$. As before, we now send $\Lambda$ to infinity after the 
redefinition 
$$\mu=\mu'\Lambda^{-1/2}\;;\qquad t=z\Lambda^{-1/2}\;.$$
 The resulting curve is 
\begin{equation}\label{Nf33}\begin{cases} v^2=z^4/4\;;\\
 w^2=\mu'^2z^2/4\;; \\
 zw=\mu'v\;, 
\end{cases} \qquad \Omega=  dvdw\frac{dz}{z}\;. \end{equation}
Imposing that the holomorphic three-form has dimension three, we find that the above curve is invariant under the $U(1)$ group 
\begin{equation}\label{nf3}\frac{2}{3}(R_{\CN=2}+I_3)\;,\end{equation}which is our candidate R-symmetry in the IR. This is precisely the combination preserved 
by the superpotential term 
$$\int d^2\theta \, \mu'\,U\;.$$ 
Notice that under (\ref{nf3}) $V=\int d^2\tilde{\theta} U$ has charge $4/3$ and $M_{SU(3)}$ has charge $2/3$. Assuming (\ref{nf3}) 
is the IR R-symmetry we can easily evaluate a and c central charges of our candidate SCFT as in the previous subsection (the central 
charges of the AD theory are given by (\ref{ccc}) setting $N=2$). The result is 
$$a=\frac{1}{6}\;;\qquad c=\frac{1}{3}\;.$$
 These are the central charges of a free theory describing eight chiral multiplets, which is 
the number of multiplets contained in $M_{SU(3)}$. Since its charge under $R_{IR}$ is precisely equal to the R-charge of 
a free chiral multiplet, we are led to the conclusion that our SCFT is simply a free theory describing a chiral multiplet in the 
adjoint of $SU(3)$ which we call again $\Psi$. 

Since in this case the theory has a nontrivial global symmetry,  it is  now possible to make  a nontrivial consistency check for our proposal by 
looking at the $SU(3)$ flavor central charge: by 't Hooft anomaly matching we know precisely the relation between the central 
charges in the UV and in the IR. Our candidate IR effective theory should then satisfy such relation. The only information we need 
for the computation is the flavor central charge of the $D_4$ AD theory which is equal to 3 (see again (\ref{ccc})). 
In the UV we have 
$$\Tr R_{IR}\,SU(3)^2=\frac{2}{3}\Tr R_{\CN=2}\,SU(3)^2=-\frac{1}{3}k_{SU(3)}=-1\;,$$
 which matches the computation in our low-energy 
effective theory 
$$\Tr R_{IR}\,SU(3)^2=\left(\frac{2}{3}-1\right)3=-1\;.$$

Let us now analyze the effect of the mass deformations in the infrared: if we slightly change the mass parameters  ($\delta m$)  in the gauge theory 
keeping them equal, the $SU(3)$ global symmetry is unbroken and from the discussion of the previous section we know that 
this deformation is described in the AD theory by the superpotential term
 $$ \int d^2\theta \, \delta m\,V\;.$$ 
 Since the R-charge of $V$ is twice 
that of $\Psi$ in the IR, we conclude that in our low-energy effective theory the deformation is described by the superpotential 
(including the $\mathcal{N}=2$ breaking term)
\begin{equation}\mathcal{W}=\delta m\,\Tr\Psi^2+\mu'\,\Tr\Psi^3\;.\end{equation}
The resulting F-term equations read
$$\label{Nf3} 3\mu'\Psi^2+2\, \delta m\Psi-\lambda{\bf 1}=0\;.$$
{\bf 1} denotes the identity matrix and $\lambda$ is a Lagrange multiplier which enforces the traceless condition. 
Modulo an $SU(3)$ transformation we can take  $\Psi$ in an  upper triangular form. Then, since all the diagonal elements satisfy 
the same quadratic equation, at least two of them are equal (for instance, $\Psi_{11}=\Psi_{22}$). If we impose that all the 
diagonal elements are equal (to zero since the matrix is traceless), one is forced to have $\lambda=0$ and all the off-diagonal elements vanishing. Otherwise, we get 
$\Psi_{11}+\Psi_{33}=-\frac{2 \delta m}{3\mu'}$. The traceless condition then implies $\Psi_{22}=\Psi_{11}=\frac{2 \delta m}{3\mu'}$, 
$\Psi_{33}=-\frac{4 \delta m}{3\mu'}$ and $\lambda=\frac{8 \, \delta m^2}{3\mu'}$. The only constraint for the off-diagonal elements is $\Psi_{12}=0$. 

In summary, we have found one vacuum in which the global symmetry is unbroken and with a mass gap,
whereas in the second group of vacua $SU(3)$ is 
spontaneously broken to (at least) $U(2)$.
 The vacuum moduli is now described by the Nambu-Gldstone modes. 
In this family of vacua one has  massless chiral multiplets. As we will explain momentarily, these findings fit well with 
the gauge theory analysis. 

From the discussion in Section 2, we also know how to deal with a generic mass deformation: the corresponding effective superpotential is 
$$ \mathcal{W}=\Tr {\cal M}  \Psi+\delta m \, \Tr\Psi^2+\mu'\Tr\Psi^3\;,$$ 
where ${\cal M} $ is a traceless diagonal matrix. Setting $\Psi=\sum_a\psi^aT_a$ this can be rewritten as 
\begin{equation}\mathcal{W}=\frac{\tilde{m}}{2}\psi_3+\frac{{\tilde n}}{2}\psi_8+\frac{\delta m}{2}\sum_a\psi_a\psi^a+\frac{\mu'}{4}\sum_{a,b,c}\psi^a\psi^b\psi^cd_{abc}\;.\end{equation}
The corresponding F-term equations can be solved exactly. It turns out that only $\psi_3$ and $\psi_8$ are nonzero, so the system 
of equations reduces to 
\begin{equation}\label{fterms}\begin{cases} 
\frac{\tilde{m}}{2}+\delta m\, \psi_3+\frac{\sqrt{3}\mu'}{2}\psi_3\psi_8=0\;;\\
\frac{\tilde n}{2}+\delta m\, \psi_8+\frac{\sqrt{3}\mu'}{4}(\psi_3^2-\psi_8^2)=0\;, 
\end{cases}\end{equation} 
which has four distinct solutions. 

This is again exactly what is expected from the original gauge theory analysis \cite{SW2}.  Let us recall that, for $N_f=3$, 
there are five singular points, three "quark" points and two dyon singularities. The AD point arises when the triplet quark 
vacuum meets one of the dyon vacua \cite{APSW}.  At a slightly off critical mass, therefore, the AD vacuum splits into the 
triplet vacua and the monopole (singlet) vacuum. In the former the global $SU(3)$ symmetry is broken to $SU(2)\times U(1)$ by a 
condensation of the triplet field.

\subsection{$SU(2)$ theory with two flavors}

After the $\CN=1$ deformation the theory can be described by the curve 
\begin{equation}\begin{cases} v^2=-\frac{\Lambda^2t^3}{(t+1)^2}\;;\\
 w^2=-\mu^2\Lambda^2(t+a)\;;\\
 w=f(t)v\;, 
\end{cases}\qquad   \Omega=dvdw\frac{dt}{t}\;. \end{equation} 
The third equation implies $a=0$ and $f(t)=\mu(t+1)/t$. In order to perform the scaling limit we consider the redefinition 
$$\mu=\mu'\Lambda^{-2/3};\quad t=z\Lambda^{-2/3}\;, $$ 
and then send $\Lambda$ to infinity. We are then left with the curve 
\begin{equation}\label{Nf2}\begin{cases} v^2=-z^3\;;\\
 w^2=-\mu'z \;;\\
 zw=\mu'v\;,
\end{cases} \qquad \Omega=dvdw\frac{dz}{z}\;, \end{equation} 
which is invariant under the $U(1)$ group (assuming as before that $\mu'$ is uncharged)
\begin{equation}\label{nf2}\frac{3}{4}R_{\CN=2}+\frac{1}{2}I_3\;, \end{equation}which is precisely the combination preserved by the $\CN=1$ deformation 
$$\int d^2\theta\mu'U\;.$$
 Under (\ref{nf2}) $U$ has charge 2, $V$ has charge 1 and $M_{SU(2)}$ has charge $1/2$. Here we come 
across a new phenomenon: (\ref{nf2}) cannot be the R-symmetry of the theory in the IR, since there is a chiral operator in the theory with 
charge smaller than $2/3$. 

We propose the following interpretation: the chiral multiplet $M_{SU(2)}$ decouples and becomes free. 
To evaluate the a and c central charges, we should then subtract the contribution from three chiral multiplets with charge $1/2$ (remember that 
$M_{SU(2)}$ transforms in the adjoint of $SU(2)$) and add the contribution of three multiplets of charge $2/3$. The final result is precisely the 
a and c central charges of three free chiral multiplets. We thus interpret the $\CN=1$ deformed AD model as a theory describing 
a chiral multiplet in the adjoint of $SU(2)$ which we call $\Psi$ as in the previous cases. 

As in the $N_f=3$ case, we can match the triangle anomaly $\Tr U(1)SU(2)SU(2)$, where $U(1)$ 
is the combination (\ref{nf2}), in the UV and in the IR: in the AD theory we find 
$$\Tr U(1)SU(2)^2=\frac{3}{4}\Tr R_{\CN=2}SU(2)^2=-1\;.$$ 
The second equality follows from the observation that $-2\Tr R_{\CN=2}SU(2)^2$ is equal to the $SU(2)$ flavor central charge which is 
$8/3$ for the $D_3$ AD theory. This is readily reproduced in our low-energy effective theory: 
$$\Tr U(1)SU(2)^2=(I_3(\Psi)/2-1)2=(1/2-1)2=-1\;.$$

Given their charges under (\ref{nf2}), we identify $V$ with $\Tr\Psi^2$ and $U$ with $(\Tr\Psi^2)^2$. 
We are then led to conclude that the effect of the mass deformations in the gauge theory 
should be accounted for at low energy by the effective superpotential 
\begin{equation}\label{pot2}\mathcal{W}= {\tilde m} \psi_{3}+\delta m\, \Tr\Psi^2+\mu'(\Tr\Psi^2)^2\;,\end{equation}
where $\psi_3$ is the diagonal component of $\Psi$, viewed as a $2\times 2$ traceless matrix. 

When ${\tilde m}=0$ the superpotential (\ref{pot2}) is $SU(2)$ preserving and acting with a global $SU(2)$ transformation we can put 
$\Psi$ in upper triangular form as before. The F-term equations become 
$$(4\mu'\Tr\Psi^2+2\, \delta m)\Psi=\lambda{\bf 1}\;,$$ whose solutions are 
$$\Psi=0\;;\qquad \Psi=\pm i\sqrt{\frac{{\delta m}}{\mu'}}\sigma_3+a(\sigma_1+i\sigma_2)\;,$$ 
where $a$ is unconstrained. 
When $a=0$ the two solutions in the second set differ by a sign so are equivalent modulo an $SU(2)$ transformation and should 
not be considered as distinct vacua. 
In the first vacuum we have a mass gap and the global symmetry is unbroken, whereas in the second group of vacua $SU(2)$ 
is broken to (at least) $U(1)$. In this class of vacua there are massless Nambu-Goldstone  modes.

If ${\tilde m}  \neq0$ the superpotential explicitly breaks $SU(2)$. Setting $\Psi=\psi^a\sigma_a$, 
we find the system of equations 
\begin{equation}\label{fterm2}\begin{cases} 
\tilde{m}\psi_1+\mu\psi_1(\sum_a\psi_a^2)=0 \;;\\
\tilde{m}\psi_2+\mu\psi_2(\sum_a\psi_a^2)=0\;; \\
\delta m +\tilde{m}\psi_3+\mu\psi_3(\sum_a\psi_a^2)=0  \;.
\end{cases}\end{equation} 
This system has three solutions: if either $\psi_1$ or $\psi_2$ is nonzero, the corresponding equation implies 
$\tilde{m}+\mu(\sum_a\psi_a^2)=0$ and the third equation reduces to $\delta m=0$, against our assumption $\delta m\neq0$. We thus find 
$\psi_1=\psi_2=0$ and $\psi_3$ is a root of the cubic equation 
$$\delta m+\tilde{m}\psi_3+\mu\psi_3^3=0\;.$$

Again, this fits correctly  with the knowledge of the  underlying gauge theory  (i.e., the AD vacuum arises as the result of a collision
of the doublet vacuum with one of the monopole/dyon vacua.)

Thus both in the $N_f= 2 $ and $N_f=3$ theories,  an interesting physics emerges when the ${\cal N}=1$ SCFT  is further deformed by 
a common noncritical quark mass, $\delta m$.  The system smoothly goes into  confining phase, with flavor symmetry spontaneously  broken. 
The free meson fields $\Psi$  make a metamorphosis into  massless Nambu-Goldstone particles.

\subsection{$SU(N)$ theory with $2N-1$ flavors \label{sec:SUN}}

By analogy with the analysis in the previous section, we expect the mass term for the adjoint multiplet $\Phi$ to be 
mapped in the SCFT to the $\CN=1$ deformation
 \begin{equation}\label{deff}\int d^2\theta\mu'U_2\;,\end{equation} 
 where $\mu'$ is 
proportional to the parameter $\mu$. It will now be seen that the analysis with the $\CN=1$ curve does confirm this expectation. 

As in the previous cases, the curve is described by a system of equations in the 
Calabi-Yau threefold parametrized (in the notation of section \ref{sec1:SUN}) by $x$, $y$ and a new coordinate $w$.
In order to determine it, in this case it is more convenient to adopt a slightly different strategy and follow the approach of 
\cite{deboer}: we exploit the fact \cite{noi} that for an $\CN=2$ gauge theory deformed by a superpotential for the chiral multiplet 
in the adjoint, the projection of the $\CN=1$ curve on the plane $(x,w)$ gives the Dijkgraaf-Vafa curve \cite{DV}. The advantage of this 
strategy is that the Dijkgraaf-Vafa curve is easy to extract from (\ref{ciao}): it is described by the equation 
$$w^2=\mu^2x(x-4\Lambda)\;.$$ 
We are then led to the system 
\begin{equation}\label{curva}\begin{cases} y^2=x^{2N-1}(x-4\Lambda)\;;\\
 w^2=\mu^2x(x-4\Lambda)\;.
\end{cases}\end{equation} 
We now perform the change of variables 
$$y=2t-x^N\;;\qquad w=2w'-\mu x\;,$$
 which brings the above system to the form 
\begin{equation}\begin{cases} t^2-tx^N+\Lambda x^{2N-1}=0\;;\\
 w'^2-\mu w'x+\mu^2\Lambda x=0\;.
\end{cases}\end{equation} 
This is the $\CN=1$ curve found in \cite{deboer}, who also observed that the holomorphic three-form is 
$$\Omega=dw'dx\frac{dt}{t}\;.$$ 
We can now notice that from (\ref{curva}) we find the relation 
$$\mu y=\pm x^{N-1}w\;.$$
Under the above change of variables this becomes 
$$2\mu t=\mu x^N\pm(2w'x^{N-1}-\mu x^N)\;.$$ 
We now choose to impose the boundary condition $$w'\rightarrow\mu x\quad \text{for} \; t\rightarrow\infty$$ 
which selects the plus sign in the previous equation. Collecting everything we find the system 
\begin{equation}\begin{cases} t^2-tx^N+\Lambda x^{2N-1}=0\;;\\
 w'^2-\mu w'x+\mu^2\Lambda x=0\;;\\
 \mu t=w'x^{N-1}\;,
\end{cases}\qquad  \Omega=dw'dx\frac{dt}{t}.\end{equation} 
The third equation encodes the information about our choice of boundary condition. 

The final step is to perform the scaling limit: making the change of variables 
$$\mu=\mu'\Lambda^{-1/2}\;;\qquad t=z\Lambda^{1/2}\;, $$
 and sending then $\Lambda$ to infinity we get the curve 
\begin{equation}\label{curvagen}\begin{cases} z^2+x^{2N-1}=0 \;;\\
 w'^2+\mu'^2x=0  \;;\\
 \mu'z=w'x^{N-1}\;,
\end{cases} \qquad  \Omega=dw'dx\frac{dz}{z}\;.\end{equation} 
We recognize in the first equation the $\CN=2$ curve for the $D_2(SU(2N-1))$ model. If we now impose the constraint $D(\Omega)=3$ 
we find that the above curve is invariant under the action of the $U(1)$ group 
\begin{equation}\label{rnew}R_{IR}=\frac{2}{3}(R_{\CN=2}+I_3)  \end{equation} 
which is precisely the combination preserved by the $\CN=1$ deformation 
$$\int d^2\theta\, \mu' \,U_2\;.$$
 Notice that this is, as expected,  the combination   found when we discussed the $SU(2)$ theory with 
three flavors. 

Using now (\ref{ccc}) it is easy to evaluate the $a$ and $c$ central charges of the resulting $\CN=1$ theory (assuming that (\ref{rnew}) 
is the infrared R-symmetry). We find 
$$a=\frac{(2N-1)^2-1}{48}\;;\qquad c=\frac{(2N-1)^2-1}{24}\;,$$ 
which are the $a$ and $c$ central charges of a free theory describing $N_f^2-1$ chiral multiplets. Since the moment map 
associated with the $SU(2N-1)$ global symmetry of the underlying $\CN=2$ theory has precisely charge $2/3$ under (\ref{rnew}), 
we conclude that the IR fixed point describes a chiral multiplet $\Psi$ in the adjoint of $SU(2N-1)$. 

The anomaly matching works in this case as well: in the $\CN=2$ theory we have 
$$\Tr R_{IR}\,SU(2N-1)^2=\frac{2}{3}\Tr R_{\CN=2}\,SU(2N-1)^2=-\frac{1}{3}k_{SU(2N-1)}=\frac{1-2N}{3}\;.$$ 
whereas  in the $\CN=1$ effective theory, 
$$\Tr R_{IR}\,SU(2N-1)^2=(R_{IR}(\Psi)-1)(2N-1)=\frac{1-2N}{3}\;.$$

Repeating the argument given for $SU(2)$ theories we conclude that a generic mass deformation in the gauge theory can be mapped in 
the IR to the following effective superpotential for the field $\Psi$:   
\begin{equation}\label{effsup} \mathcal{W}(\Psi)=\frac{1}{2}\sum_i {\tilde m}_i\Psi_i+ \delta m\, \Tr\Psi^2+\mu'  \Tr \Psi^3,\end{equation} 
where ${\tilde m}_i$ are the $SU(2N-1)$ mass parameters and $\Psi_i$ are the components of $\Psi$ along the Cartan subalgebra. 

At the AD point, i.e.,  without mass deformation,  ${\tilde m}_{i}=\delta m=0$,   one finds the equation for $\Psi$ 
\begin{equation} 3 \mu^{'} \Psi^{2}  -  \lambda  {\bf 1}=0
\;, \qquad   \Tr \, \Psi=0\;,\label{psieq}
\end{equation}
where $\lambda $ is the Lagrange multiplier.   By a unitary transformation $\Psi$ can be  always put into upper triangular form (Schur's decomposition), 
\begin{equation}\Psi =   \left(\begin{array}{cccc}a_1 & * & \cdots & * \\0 & a_2 & \ddots & * \\\vdots & \ddots & \ddots & * \\0 & \cdots & 0 & a_{N_f}\end{array}\right)\;.
\end{equation}
The first of the equations (\ref{psieq}) tells that the diagonal elements satisfy
\begin{equation}3 \mu^{'} a_{i}^{2} =\lambda\;, \qquad  \text{i.e.} \qquad  a_{i}=\pm  \sqrt{\lambda /  3 \mu^{'} }\;. 
 \end{equation}
 As $N_{f}=2N-1$ is odd, after taking into account the second equation one has necessarily 
 \begin{equation}    \lambda=0\;,  \qquad a_{i}=0\;,   \quad  \forall i\;. 
 \end{equation}\;
 Equation (\ref{psieq}) then reduces to
 \begin{equation} \Psi^{2}=0\;. \label{Higgsbr}
 \end{equation}
 This vacuum moduli represents various Higgs branches emanating from the AD point.  It will be seen in the next section that this result 
 reproduces the structure of the vacuum moduli space of the gauge theory.
 
 Let us analyze now the effect of mass deformation,  $\delta m \ne  0$ ($\delta m$ is the shift of the common mass away  from the critical mass 
 $m^{*}= \Lambda/N$).   The equations for $\Psi$ are now  
 \begin{equation} 2 \,\delta m \Psi +  3 \mu^{'} \Psi^{2}  -  \lambda  {\bf 1}=0\;, \qquad   \Tr \, \Psi=0\;. \label{psieqbis}
\end{equation}
The diagonal elements  $  a_{i}$ are solutions of the second-order equation
\begin{equation}  2 \,\delta m \,a +  3 \mu^{'} a^{2}  -  \lambda  =0\;. 
\end{equation}
Denoting the two solutions as 
\begin{equation} a_{\pm} =  \frac{1}{3\mu^{'}} \{  - \delta m \pm \sqrt{ (\delta m)^{2} +  3\mu^{'} \lambda} \}\;, 
\end{equation}
and assuming that $r$ of the diagonal elements are equal to $a_{+}$, the remaining $N_{f}-r$ to $a_{-}$ (we call such a solution 
$r$ vacuum), 
 the equation  $\Tr \, \Psi=0$  gives 
\begin{equation}   r\, a_{+} + (N_{f}-r) \, a_{-}  =0\;,  \qquad r=0,1,\ldots,  [\frac{N_{f}}{2} ]\;,
\end{equation}
which determines $\lambda$ for each $r$ vacuum. The generic solution has the form 
\begin{equation} \Psi=\left(\begin{array}{c|c} 
a_{+}{\bf 1}_{r} & * \\
\hline 
0 & a_{-}{\bf 1}_{N_f-r}\end{array}\right),\end{equation}
where the elements in the upper right block are unconstrained.
 In a $r$ vacuum,  the global $SU(N_{f})\times U(1)$ symmetry is broken to  $U(r)\times U(N_{f}-r)$.  
The associated $2r (N_{f}-r)$   Nambu-Goldstone modes arise from the Higgs branches around the AD point. 

Finally, when the mass parameters are generic, the F-term equations become 
\begin{equation}\label{psigen}  3\mu^{'} \Psi^{2}+2 \, \delta m \Psi+{\cal M} =0\;,\end{equation}
where ${\cal M}$ is a diagonal matrix whose eigenvalues all have multiplicity one (this is what is meant 
by generic). It is possible to show that under this assumption all the solutions of (\ref{psigen}) are diagonal and since every 
element on the diagonal satisfies a quadratic equation, one finds $2^{N_f-1}$ solutions.    

 We will see in the next section that this result agrees with the counting of 
vacua in the underlying $SU(N)$  gauge theory.  
 In the next section, we shall  study in more detail the properties of our critical point from the original gauge theory 
 perspective. 

\section{The gauge theory perspective  \label{sec:GTP}}   

 The softly broken 
${\cal N}=2$  $SU(N)$ gauge theory (we shall limit ourselves to the particular odd flavor case,  $N_f= 2 N-1$ for concreteness)  is described by the  superpotential, 
\begin{equation}\label{trlevel}\mathcal{W}=\sqrt{2}\, \widetilde{Q}^i\Phi Q_i+m_{i}\, \widetilde{Q}^iQ_i+\frac{\mu}{2}\Tr\Phi^2.\end{equation}
 Using the generalized Konishi anomalies \cite{CDSW, CSW}
 all the chiral condensates of the theory can be expressed in terms of the parameters of the theory, the gluino condensate and
the trace of the matrix of meson vevs. The latter two quantities can in turn be computed by extremizing the Dijkgraaf-Vafa superpotential. 
The corresponding equations are \cite{iolorenzo}: 
\begin{equation}\label{trace}
\frac{\mu}{2}\left[(N_{f}-2r)\sqrt{(a-\eta)^{2}-4S/\mu}-(2N-N_f)a-N_{f}\eta\right]=0\;,
\end{equation}
\begin{equation}\label{gluino}
(N_{f}-2r)\log\left(\frac{a-\eta}{2}-\frac{1}{2}\sqrt{(a-\eta)^{2}-\frac{4S}{\mu}}\right) + \log\left(\frac{\mu^{N-r}\Lambda^{2N-N_f}}{S^{N-r}}\sqrt{2}^{N_f}\right)=0\;. 
\end{equation}
where 
$\eta$ is the bare quark  mass  
(we consider here only the case with degenerate mass parameters for the flavors)
$$\eta\equiv-\frac{m}{\sqrt{2}}\;,$$
and  
$$ a \equiv \frac{\sqrt{2}}{N\mu}\langle\tilde{Q}^{i}Q_{i}\rangle    $$
is the meson condensate.  
The parameter $r$ labels different vacua and runs from 0 to $[N_f/2]$ (see \cite{iolorenzo} for details). The SW curve of the 
gauge theory in a $r$ vacuum can be written in the form 
$$y^2=(x-\eta)^{2r}Q_{N-r-1}^2(x)(x-a+2\sqrt{S/\mu})(x-a-2\sqrt{S/\mu})\;.$$ 
We obtain the point of interest for us when one of the two unpaired roots in the previous equation coincides with $\eta$ (see 
section 2). This condition imposes the constraint 
\begin{equation}\label{simple}(a-\eta)^{2}-4S/\mu=0\;,\end{equation} 
which greatly simplifies equations (\ref{trace}) and (\ref{gluino}) since all the terms inside the 
square roots are set to zero. From (\ref{trace}) and (\ref{simple}) we then immediately find 
\begin{equation}\label{simple2}a=-\frac{N_f}{2N-N_f}\eta
\;;\qquad S=\mu\left(\frac{2N}{2N-N_f}\right)^2\frac{\eta^2}{4}\;.    \end{equation} 
Plugging these two relations in (\ref{gluino}) we find an equation for $\eta$ whose solutions are 
\begin{equation}\label{massaad}\eta^{*  }=-\frac{m^{*}}{\sqrt{2}}  =\omega_k\frac{2N-N_f}{N}2^{N_f/(4N-2N_f)}\Lambda\;,\end{equation} 
where 
$$\omega_k=e^{\pi i(2k+1)/(2N-N_f)}\;, \qquad k=0,\dots,2N-N_f-1\;.$$ 
Notice that in this solution the dependence on $r$ disappears, implying that the various $r$ vacua (which are distinct for a generic 
value of the mass parameter) merge together for the above choices of $\eta$. We can now recover the vacuum counting of section 3: 
when we change the value of the common bare mass our singular point splits into $[N_f/2]$ $r$ vacua (as we will see later there are 
flat directions emanating from these points). If now  the bare masses are taken to be generic each $r$ vacuum further splits into 
$\binom{N_f}{r}$ vacua \cite{Carlino:2000uk}. We thus find a total of 
$$\sum_{r=0}^{[N_f/2]}\binom{N_f}{r}=2^{N_f-1}$$ vacua, in agreement with the result of the previous section.

In order to compare the results with the analysis of Section~\ref{sec:SUN}, we shall set $N_{f}=2N-1$ below, which will make the formulas  somewhat simpler.  

We shall now see that for the above choice of the mass parameter $\CN=2$ SQCD deformed by a mass term for the chiral multiplet in the adjoint has a nontrivial moduli space, with the same structure as that found in Section~\ref{sec:SUN}. 

Since $\Phi$ enters quadratically, we can integrate it out using the equations of motion and work in terms of the meson field only. 
Usually in the literature the resulting equations are solved by taking the meson matrix to be diagonal. Indeed, when the 
mass parameters are generic this is not restrictive: if we consider the variations 
$$\delta\widetilde{Q}^j=\widetilde{Q}^i;\quad \delta Q_i=Q_j \quad (i\neq j)$$ we find the relations 
$$\delta\widetilde{Q}^j\frac{\partial\mathcal{W}}{\partial\widetilde{Q}^j}=\sqrt{2}\widetilde{Q}^i\Phi Q_j+m_j\widetilde{Q}^iQ_j=0\;,$$ 
$$\frac{\partial\mathcal{W}}{\partial Q_i}\delta Q_i=\sqrt{2}\widetilde{Q}^i\Phi Q_j+m_i\widetilde{Q}^iQ_j=0\;.$$ 
These equations are not subject to quantum corrections associated with anomalies simply because the anomaly is trivial for these 
transformations. Taking the difference of the two equations we get 
$$(m_j-m_i)\widetilde{Q}^iQ_j=0\;,$$ 
and if $m_i\neq m_j$ for all $i\neq j$ we directly conclude that the matrix of meson vevs is diagonal. However, since we are interested 
in the case of equal mass parameters, this argument does not apply. 

Let us integrate out the multiplet in the adjoint and rewrite the superpotential in terms of the meson field only. From the equations 
of motion we find 
\begin{equation}\label{adjoint}\Phi_a=-\frac{2\sqrt{2}}{\mu}\widetilde{Q}^iT_aQ_i \;, \end{equation}
and plugging this back in (\ref{trlevel}) we get 
$$\mathcal{W}=-\frac{1}{\mu}\left(\Tr M^2-\frac{1}{N}(\Tr M)^2\right)+\Tr mM\,.$$ 
The space of vacua is obtained by extremizing the above superpotential supplemented with the suitable nonperturbative 
superpotential term. The matrix of meson vevs thus satisfies the equation 
\begin{equation}\label{supnp}-\frac{2}{\mu}M^2_{ij}+\left(\frac{2\Tr M}{\mu N}+m\right)M_{ij}+(M\partial\mathcal{W}_{NP}/\partial M)_{ij}=0\;.\end{equation} 
The actual form of $\mathcal{W}_{NP}$ depends on the value of $N_f$. Instead of discussing the various cases separately, we can proceed as follows: 
considering again the variation 
$$\delta Q_i=Q_j\;,$$
 we find the equation 
$$\sqrt{2}\widetilde{Q}^i\Phi Q_j+m_i\, \widetilde{Q}^iQ_j-S\delta^i_j=0\;,$$
 where $S$ is the gluino condensate. The last term arises 
as a consequence of the Konishi anomaly. Plugging now in the above equation (\ref{adjoint}) we find 
\begin{equation}\label{meseq}-\frac{2}{\mu}M^2_{ij}+\left(\frac{2\Tr M}{\mu N}+m\right)M_{ij}-S\delta_{ij}=0  \;,
\end{equation}
and by comparing this with (\ref{supnp}) we obtain the relation $$S\delta^i_j=-(M\partial\mathcal{W}_{NP}/\partial M)_{ij}\;.$$
The advantage of this observation is that by setting the mass parameter $m$ to the  critical value $m=m^{*}$ (\ref{massaad}) we select the AD point 
we are interested in and we know already the value of the gluino condensate (and of $\Tr M$) there (see Eq~(\ref{simple2})). 

Using (\ref{simple}) and (\ref{simple2}) we can rewrite (\ref{meseq}) in the form 
\begin{equation} M^2_{ij}-2\alpha M_{ij}+\alpha^2\delta_{ij}=0\;;  \qquad \alpha\equiv\frac{N\mu m^{*}}{4N-2N_f}=\frac{\Tr M}{N_f}\;.    \label{critical}
\end{equation} 
  The above equation can be  recast into the form, 
  \begin{equation}\widetilde{M}^2=0\;,    \label{M2}  \end{equation} 
  where ${\tilde M}$ denotes  the traceless part of $M$: 
\begin{equation}\widetilde{M}\equiv  M-\alpha I \;. \end{equation}
Notice that this precisely reproduces the result (\ref{Higgsbr}) obtained in our effective theory : 
 $\Psi^2=0$,   
 
When the quark masses are taken slightly off the critical value,  one still has 
equation  (\ref{meseq}),  but  $\Tr M$ and  $S$ are shifted according to Eqs.~(\ref{simple2}), and consequently   Eq.~(\ref{meseq}) no longer reduces to  Eq.~(\ref{M2}).
In order to see how the latter gets modified, let us set   ($\eta\equiv  -m/\sqrt{2}$)
\begin{equation} \eta=  \eta^{*}+ \epsilon\,, \qquad ( \eta^{*}= - {m^{*}}/{\sqrt{2}}\;; \quad  \epsilon = - {\delta  m}/{\sqrt{2}})\;; 
\label{devia1}\end{equation}
\begin{equation} a= - N_{f} ( \eta^{*}+ \alpha \epsilon + \gamma \epsilon^{2} ) +  O(\epsilon^3) \;, \label{devia2} 
\end{equation}
and 
\begin{equation} S =  \mu (2N)^{2}\frac{(\eta^{*})^{2}}{4}  + \beta \epsilon + \delta \epsilon^{2}+  O(\epsilon^3)  \;.  \label{devia3} 
\end{equation}
in  Eq.~(\ref{trace}) and Eq.~(\ref{gluino}). Apparently it is a straightforward exercise to solve them order by order in  $\epsilon$, and to insert the solutions into Eq.~(\ref{meseq})
to find the modification of the equation for $\tilde M$. On general grounds one expects  (\ref{M2})  to be modified to the form, 
\begin{equation}  {\tilde M}^2  +   C_1 {\tilde M} + C_2 \,{\bf 1} =0\;,   \label{Meqn}
\end{equation}
with
\begin{equation}  C_1= O(\epsilon)\;, \qquad  C_2= O(\epsilon)\;,\;  
\end{equation}
which would appear to reproduce nicely equation (\ref{psieqbis})   for the  $\Psi$  fields found in the previous section.  

Actually it requires a  more careful analysis to show that equations (\ref{Meqn}) and (\ref{psieqbis}) have indeed the same structure.  
The problem is that  Eq.~(\ref{trace}) and Eq.~(\ref{gluino})  clearly depend on $r$, and it is natural to expect that  the solutions for   $\alpha, \beta, \gamma, \delta, \ldots $  depend 
on it,  even though  to the zeroth order in $\epsilon$ (i.e., at exactly the AD point) the solution for $a=\Tr M$ and $S$ were found not to depend on $r$  (the coalescent $r$ vacua).  As a consequence it is natural to expect that the coefficients $C_1$, $C_2$ would depend on $r$. 
On the other hand, the coefficients of the terms $\Psi^1$,  $\delta m \sim \epsilon$,   and $\Psi^0$,  $\lambda$  in (\ref{psieqbis}) are clearly independent of $r$, {\it a priori}.   $\lambda$ is just a
Lagrange multiplier. However, the {\it  solutions} of  (\ref{psieqbis}) are classified by an integer $r=0,1,\ldots, [N_f/2]$:   the resulting $\lambda $ does depend on the solution, hence on $r$. 
How could such a system of equations be consistent with Eq.~(\ref{meseq}),  with  (\ref{devia1}), (\ref{devia2}), (\ref{devia3})?   

In spite of the apparent $r$ dependence of the equations, quite surprisingly the solution for the first order shifts $\alpha$ and $\beta$  from  Eq.~(\ref{trace}) and Eq.~(\ref{gluino}) 
turn out to be independent of $r$ and given simply by   (for $N_f= 2N-1$)
\begin{equation}   \alpha = \frac{N-1}{2N-1}\;,  \qquad   \beta=  N^2  \mu  \eta^*\,,  \label{firstorder}
\end{equation}
and accordingly, 
\begin{equation} C_1=  \frac{\mu}{\sqrt {2}  (2N-1)} \epsilon =  \frac{\mu}{\sqrt {2} N_f}  \epsilon \;.    \label{C1}
\end{equation}

By using the first order results, (\ref{firstorder}), one then readily shows that the order $\epsilon$ contribution to $C_2$ exactly cancels, so that $C_2 = O(\epsilon^2)$.  In order to find 
$C_2$,  the second order deviations of  $\Tr M$ and $S$ from the critical values, $\gamma$ and $\delta$,  must be calculated.  Now a second surprise is that actually
$\gamma$ and $\delta$ turn out to be indeterminate.  This can be seen from the fact that by using
\begin{equation} \sqrt{(a-\eta)^{2}-4S/\mu} =  \frac{2 N- N_f}{N_f - 2r}  a +   \frac{N_f }{N_f - 2r}    \label{tobeable} \eta 
\end{equation}
following from Eq.~(\ref{trace})  in  Eq.~(\ref{gluino}), there is actually only one equation  \footnote{For the first order shifts $\alpha, \beta$, there is one more condition 
that the argument of the square root must not contain   the order $\epsilon$ terms, i.e., that the first equation (\ref{tobeable}) be compatible with  expansion in powers of $\epsilon$. }   for  $\gamma$ and $\delta$.   Therefore one finds that
\begin{equation}  C_2 = O(\epsilon^2)\;,      \label{C2}
\end{equation}
with indeterminate coefficient. This is what allows Eq.~(\ref{Meqn}) to possess solutions \footnote{If $C_{2}$ were 
{\it a priori} fixed, equation (\ref{Meqn}) would not have nontrivial solutions at all,  in general.  
} corresponding to various $r$ vacua.   $C_{2}$ is determined for each solution (and is indeed of the order $O(\epsilon^{2}$)).

The conclusion is that the equation for ${\tilde M}$, with   (\ref{Meqn}), (\ref{C1}), (\ref{C2}),      
has  the same structure as that for $\Psi$,   Eq.~(\ref{psieqbis}).    The degrees of freedom of the system are
the massless fields 
 describing the traceless part of the  gauge-invariant  mesons, ${\tilde M}$. 

\section{Discussion}

\begin{figure}[h!t]
\epsfxsize=8 cm
\centerline{\epsfbox{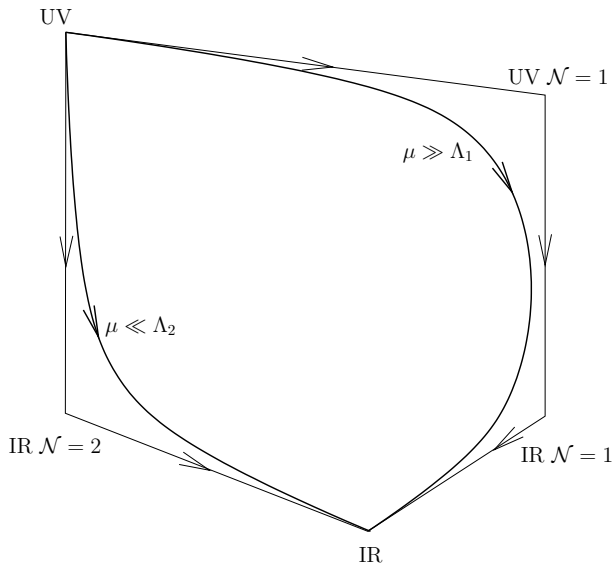}}
\caption{{\footnotesize  RG flow for various values of $\mu$.}}
\label{flowsusy}
\end{figure}

The analysis of Section 3 is done by first flowing into the AD infrared fixed point SCFT of ${\cal N}=2$ theories, and then upon  ${\cal N}=1$ perturbation, 
 \begin{equation}   \mu \, \Tr \, \Phi^{2} \;,  
 \end{equation}
by following the RG path further down to the ${\cal N}=1$ SCFT.  This corresponds to taking 
\begin{equation}   \mu \ll   \Lambda_2\;,
\end{equation}
  (the left path of Fig.~\ref{flowsusy}).
Remarkably, while the  ${\cal N}=2$  AD points are interacting  SCFT with relatively nonlocal 
massless fields, the corresponding  ${\cal N}=1$ SCFTs  (physics below the mass scale $\mu$)  are found to be described   by a free massless chiral meson supermultiplet  $\Psi$  in the adjoint representation of the global $SU(N_{f})$, in all cases studied.  

  This way, the moduli space of vacua (Higgs branches) 
emanating from the origin $\Psi=0$ at exactly the ${\cal N}=1$ AD point have been exhibited. 
Furthermore, the deformation by noncritical quark masses  was studied.   In particular,  for noncritical but common quark masses, it was found that  the vacuum splits into a set of $r$ vacua, with the global symmetries broken  in each of them as
\begin{equation}    U(N_{f}) \to U(r)\times U(N_{f}-r)\;, 
\end{equation}
the resulting Nambu-Goldstone modes (massless moduli) descend from the Higgs branches emanating from the origin at the AD point. The equation for the Higgs branch $\Psi^2=0$ (\ref{Higgsbr}) contains various sub-branches of rank $r$ corresponding to the various Higgs branches emanating from the $r$-vacua colliding  at the AD point.\footnote{The maximal complex dimension of the Higgs branch is $2 r(N_f -r)$ with $r = [N_f/2]$. At the origin of the Higgs branch we found that the number of degrees of freedom is enhanced to $N_f^2 -1$. This is consistent with the vanishing of the dual-quarks condensate in the AD point \cite{Bolognesi:2008sw}.}  

As a consistency check,  these results have been reproduced in Section \ref{sec:GTP} directly from the original ${\cal N}=2$  gauge theory - softly broken to ${\cal N}=1$ -  by use of the Dijkgraaf-Vafa superpotential and Konishi anomalies, by appropriate identification of the relevant ${\cal N}=1$ vacua. 
Such an agreement shows that the ultimate destination of  the RG flow into the infrared is the same, regardless of which routes
are taken to reach it, see  Fig.~\ref{flowsusy}. 

As a still further check,  we have re-derived the results of Section~\ref{sec:GTP}  (hence of Section 3) by using the effective action description
of the softly broken ${\cal N}=2$  SQCD,  in terms of the mesons, baryons and (for $N_{f}>N$)  dual quarks  (see Appendix A).
 Even though the analysis of the Appendix A  is equivalent to the one in Section~4, it has the advantage of illustrating the fact that the meson field $M$ appearing in Seiberg's dual theory (the first two terms of Eq.~(\ref{Seiberg})) and the  fields  in Eq.~(\ref{critical}) and Eq.~(\ref{M2}) are the same degrees of freedom.

Thus  the results of the analysis on the ${\cal N}=1$  deformation of the ${\cal N}=2$  Argyles-Douglas 
theories proposed in Sections \ref{sec:Singular} and \ref{sec:N=1},  based on the recent findings  on the ${\cal N}=1$  curves,   are consistent with all known facts about 
softly-broken  ${\cal N}=2$   SQCD. In particular, the massless mesons in the adjoint representation of the global symmetry group found there, and the meson fields appearing in Seiberg's dual SQCD, are closely related:  they can be regarded as the same objects, seen through different RG  flow paths.  

  The results of the analysis in Appendix A furthermore suggest that our conclusion is generally valid for $SU(N)$ theories, 
  both for odd and even  $N_{f}$.  For $SU(2)$ theories this has been shown for $N_{f}=1,2,$ and $3$ by our new method in 
  Section~\ref{sec:N=1}. But for $N\ge 3$ the analysis of Section~\ref{sec:N=1} works only for $N_{f}$ odd. 
  For theories with even $N_{f}$ the ${\cal N}=2$ vacua of interest are described by the GST duals \cite{GST}, which are 
  nonlocal theories involving an infrared free $SU(2)$ gauge group. The method described in Section~\ref{sec:N=1} of perturbing 
  the ${\cal N}=2$ SCFT with an adjoint mass term and flowing further down to the IR does not seem to apply straightforwardly 
  in this case. It is left as a future problem to generalize appropriately the analysis of Section~\ref{sec:SUN} to general $N_{f}$.


To summarize, the class of  ${\cal N}=2$  AD systems  (strongly-interacting theories, described by relatively nonlocal quarks and monopoles),  
when perturbed  by a relevant ${\cal N}=1$  deformation, flow in the infrared to a local system  described by a free gauge-invariant meson in the adjoint representation of the flavor symmetry group.  Such an infrared physics is somewhat reminiscent of the low-energy pion physics of QCD, even though  the ${\cal N}=1$  AD system is a conformal fixed-point theory.   
Perhaps, an even stronger analogy can be drawn by considering, as we did,  the systems slightly off the AD points (by noncritical quark masses $m^*+\delta m$).  The low-energy degrees    
of freedom in these {\it confining}  vacua are the Nambu-Goldstone bosons of flavor symmetry breaking, which are a disguise of the free mesons  defining the nearby  infrared fixed-points. 

\section*{Acknowledgments} 
This work was partially supported by the ERC Advanced Grant "SyDuGraM", by FNRS-Belgium (convention FRFC PDR T.1025.14 and 
convention IISN 4.4514.08) and by the ``Communaut\'e Fran\c{c}aise de Belgique" through the ARC program.  The work by S.B. and K.K. is supported by the INFN
special research project, ``GAST''  (Gauge and String Theries).

\appendix

\section{Softly broken ${\cal N}=1$ SQCD and Seiberg duality}

In this Appendix the analysis of Section 3 is repeated by making use of the familiar infrared phases  of ${\cal N}=1$ SQCD,   including 
Seiberg's dual SQCD and superconformal fixed points,  the low-energy physics in free magnetic phase,  the infrared-free $r$-vacua, etc  \cite{Sei}, \cite{Carlino:2000uk}.  The relation between the free ``mesons''  in  the adjoint representation of the  global symmetry group found in Sections~\ref{sec:Singular} and \ref{sec:N=1} and  the meson field appearing in Seiberg's dual $SU(N_{f}-N)$ theory become transparent. The discussion below closely follows that of \cite{Carlino:2000uk}.

The relation between the ${\cal N}=2$ and ${\cal N}=1$ dynamical scales is given by the relation
\begin{equation}
\label{scalerelation}
\mu^{N} \Lambda_2^{2N -N_f} = \Lambda_1^{3N -N_f}\;.
\end{equation}
In the analyses of Sections ~\ref{sec:Singular} and  \ref{sec:N=1},  the adjoint scalar mass is chosen such that  $\mu \ll \Lambda_2$:  in this case  the RG flow passes very close to the ${\cal N}=2$ IR fixed point, and  the analysis made in  those  Sections   is appropriate. 

Let us choose  $\mu \gg \Lambda_1$  this time. The RG  flow passes now very close to  
various  ${\cal N}=1$ fixed points. 
See Figure \ref{flowsusy}. 
We shall see that in the far IR, the end point of all RG flows is always the same.

One must distinguish various IR phases of ${\cal N}=1$ SQCD. 
For $N_f < N$ the low-energy physics is described by  a mesonic field $M= \tilde{Q} Q$ plus the  Affleck-Dine-Seiberg (ADS) instanton  superpotential. 
The total superpotential, after integrating  out the adjoint field $\Phi$,  is then 
\begin{equation}
\label{superpot}
W_{eff}  = m \, {\tr} M -\frac{1}{ \mu} \left(\tr M^2 - \frac{1}{N} (\tr M)^2\right)
+(N-N_f)\frac{\Lambda_1^{\frac{3N-N_f}{N-N_f}}}{ (\det M)^{\frac{1}{N-N_f}}}\;.
\end{equation}
It is convenient to separate the identity component of the meson $M$ from the traceless part
\begin{equation}
M = v {\bf 1}_{N_f} + \tM \;;\qquad \quad \tr \tM = 0\;.
\end{equation}
By expanding the effective action to second order in $\tM$,  one finds the equations determining  the vacua
\bea
\label{first}
\partial_v W_{eff} &=&  m\,  N_f  - \frac{2}{ \mu}  \frac{N_f(N-N_f)}{N} v 
- N_f \frac{\Lambda_1^{\frac{3N-N_f}{N-N_f}}} {v^{\frac{N}{N-N_f}}} =0\;; \\
\partial_{\tM} W_{eff} &=&  -\frac{2}{ \mu} \tM + \frac{\Lambda_1^{\frac{3N-N_f}{N-N_f}}}{v^{\frac{2N - N_f}{N-N_f}}}\tM =0\;.
\eea
We are interested in finding a vacuum with massless meson field. This leads to the vev of the trace part of $M$
\begin{equation}
v^{\frac{2N - N_f}{N-N_f}} =  \frac{\mu \Lambda_1^{\frac{3N-N_f}{N - N_f}} }{2}\;,
\end{equation}
and to the determination of the critical mass
\begin{equation}
\label{result}
 m =   m^{*}=\frac{2^{\frac{N}{2N - N_f}}(2N - N_f)}{N} \frac{\Lambda_1^{\frac{3N-N_f}{2N-N_f}}}{\mu^{\frac{N}{2N-N_f}}} \;.
\end{equation}
This, together with  (\ref{scalerelation}), leads to  the same result as found in Section~\ref{sec:SUN} and in Section~\ref{sec:GTP}. 

For $N_f = N$ the low energy of ${\cal N}=1$ SQCD is descried by the meson plus the singlet baryons $B$ and $\tilde{B}$. The superpotential is 
\bea
W_{eff}  = m \ \tr M -\frac{1}{ \mu} \left(\tr M^2 - \frac{1}{N} (\tr M)^2\right)
+ \lambda  \left( \det M  - \tilde{B}  B -\Lambda_1^{2N} \right)\;,
\eea
where $\lambda$ is a Lagrange multiplier that enforces the condition $\det M  - \tilde{B}  B -\Lambda_1^{2N} = 0$. As before we  expand around $\tM = 0$  to $O(\tM ^{2})$;    
the vacuum equations are
\bea
\label{first1}
\partial_v W_{eff} &=&  m\,  N  +\lambda N v^{N-1}\;; \\
\partial_{\tM} W_{eff} &=&  -\frac{2}{ \mu} \tM - \lambda v^{N-2} \tM\;;  \\
\partial_{\lambda} W_{eff} &=& v^N +\frac{1}{2}v^{N-2} \tr \tM^2 - \tilde{B}  B -\Lambda_1^{2N} \;.
\eea
These equations are solved by $\tM=0$. Enforcing the condition that the traceless part of the meson is massless one has  $ \lambda = - \frac{2}{\mu \Lambda_1^{2N-4}}$. This, together with $B=\tilde{B}=0$ and $v=\Lambda_1^{2}$,  gives the value of the mass
\begin{equation}
 m =  \frac{2\Lambda_{1}^2}{\mu}\;,
\end{equation}
which again  agrees with  (\ref{result})   for  $N_f =N$.

For $N_f = N + 1$ the baryons carry  flavor charges and the superpotential is 
\begin{equation}
W_{eff}  = m  \, \tr M -\frac{1}{ \mu} \left(\tr M^2 - \frac{1}{N} (\tr M)^2\right)
- \frac{1}{\Lambda_1^{2N-1}} \left( \det M  - \tilde{B} M B \right)\;.     
\end{equation}
In the vaccum where $M$ condenses to a diagonal form, the baryons become massive and can be integrated-out. One is then left with the same effective superpotental (\ref{superpot}).

Finally consider  the cases $N_f > N+1$.  This is the case of interest for direct comparison with the results of Section~\ref{sec:SUN} and Section~\ref{sec:GTP}.
The IR theory is a $SU({\tilde N})= SU(N_f-N)$ gauge theory described by  dual quarks and a meson field $M$ \cite{Sei}.
The superpotential is
\begin{equation}
W = \tilde{q} M q + m \tr M -\frac{1}{ \mu} \left(\tr M^2 - \frac{1}{2} (\tr M)^2\right)  \;.\label{Seiberg}
\end{equation}
The F-term equations for the dual quarks  are
\begin{equation}
M \,q = 0 \;;  \qquad \tq \,M = 0 \;, 
\end{equation}
which means that meson  and dual quark vevs are orthogonal in flavor.  The vacua with nonvanishing dual quark  vevs are found by solving straightforwardly \cite{Carlino:2000uk}
the meson equation of motion following from Eq.~(\ref{Seiberg}). 

 The vacuum we are looking for turns out to correspond  to the one in which the vev of the meson field has the maximum rank,  $N_{f}$. 
 The dual quarks are all massive, and after integrating them out, one gets an effective superpotential \cite{Carlino:2000uk},  
 \begin{equation}
W_{eff}  = m \ {\rm Tr} M -\frac{1}{2 \mu} \left({\tr  M}^2 - \frac{1}{N} ({\tr M})^2\right)
+\frac{1}{\Lambda_1^{\frac{3N-N_f}{N_f-N}}} (\det M)^{\frac{1}{N_f-N}}\;,
\end{equation}
 the last term having the same form as the instanton superpotential (\ref{superpot}), or better, its  analytic continuation to $N_{f}> N$.
 In this vacuum the dual quarks are massive, the  (dual) gauge sector has a mass gap, and the trace part of the meson field $v$ is also massive. The only degrees of freedom in the far IR are the traceless mesons.  The analysis from this point on has already been worked out in Section~\ref{sec:GTP}.



\end{document}